\documentclass[11pt]{article}
\pdfoutput=1

\usepackage[T1]{fontenc}
\usepackage[utf8]{inputenc}
\usepackage{microtype}
\usepackage{xcolor,comment}
\definecolor{tabblue}{HTML}{1F77B4}  
\definecolor{taborange}{HTML}{FF7F0E} 
\colorlet{tabblueDark}{tabblue!70!black}
\colorlet{taborangeDark}{taborange!70!black}
\usepackage{eso-pic}

\usepackage{amsmath,amssymb,mathtools}
\allowdisplaybreaks
\usepackage{tensor}
\usepackage{bbm}
\usepackage{braket}

\usepackage{fullpage}
\usepackage[colorlinks=true,
            linkcolor=tabblueDark,
            citecolor=tabblueDark,
            urlcolor=tabblueDark]{hyperref}
\usepackage{cleveref} 

\usepackage{graphicx}
\usepackage{subcaption}

\graphicspath{{img/}}

\usepackage{cite}

\usepackage{authblk}

\newcommand{\beq}{\begin{equation}}
\newcommand{\eeq}{\end{equation}}

\newcommand{\cl}[1]{\mathcal{#1}}
\renewcommand{\[}{\begin{equation}\begin{aligned}}
\renewcommand{\]}{\end{aligned}\end{equation}}

\newcommand{\preprintnum}{\footnotesize CERN-TH-2025-166, ADP-25-30/T1292}

\numberwithin{equation}{section}

\title{\bf Spin versus Magic: Lessons from Gluon and Graviton Scattering}

\author[1]{John Gargalionis\thanks{\href{mailto:john.gargalionis@adelaide.edu.au}{john.gargalionis@adelaide.edu.au}}}
\author[2]{Nathan Moynihan\thanks{\href{mailto:n.moynihan@qmul.ac.uk}{n.moynihan@qmul.ac.uk}}}
\author[3,4]{Sokratis Trifinopoulos\thanks{\href{mailto:sokratis.trifinopoulos@cern.ch}{sokratis.trifinopoulos@cern.ch}}}
\author[1]{Ewan N.\ V.\ Wallace\thanks{\href{mailto:ewan.n.wallace@adelaide.edu.au}{ewan.n.wallace@adelaide.edu.au}}}
\author[2]{Chris D.\ White\thanks{\href{mailto:christopher.white@qmul.ac.uk}{christopher.white@qmul.ac.uk}}}
\author[1]{Martin J.\ White\thanks{\href{mailto:martin.white@adelaide.edu.au}{martin.white@adelaide.edu.au}}}

\affil[1]{\it ARC Centre of Excellence for Dark Matter Particle Physics \& CSSM, Department of Physics, University of Adelaide, Adelaide, SA 5005, Australia}
\affil[2]{\it Centre for Theoretical Physics, School of Physical and Chemical Sciences, Queen Mary University of London, 327 Mile End Road, London E1 4NS, UK}
\affil[3]{\it Theoretical Physics Department, CERN, 1211 Geneva 23, Switzerland}
\affil[4]{\it Physik-Institut, Universit\"at Z\"urich, 8057 Z\"urich, Switzerland}

\date{{}}

\begin{document}

\AddToShipoutPictureFG*{%
  \put(\LenToUnit{\paperwidth-1in},\LenToUnit{\paperheight-0.8in}){%
    \makebox(0,0)[tr]{\preprintnum}%
  }%
}

\maketitle

\begin{abstract} 
  The quantum property of {\it non-stabiliserness}, also known as
  {\it magic}, plays a key role in designing quantum computing systems. How to
  produce, manipulate and enhance magic remains mysterious, such that concrete
  examples of physical systems that manifest magic behaviour are sought after. In
  this paper, we study two-particle scattering of gluons and gravitons in
  Yang--Mills theory and General Relativity, as well as their supersymmetric extensions. This provides an
  interesting case of two-qubit systems, differing only in the physical
  spin of the qubits. We show that magic is generically produced in both theories,
  and also show that magic typically decreases as the spin of the qubits
  increases. The maximal magic in each case is found to be substantially less than
  the known upper bound. Differences in the profile of magic generation can be
  traced to the known physics of each theory, as manifested in relations
  between their respective scattering amplitudes. Our case study may provide
  useful insights into understanding magic in other systems. 
\end{abstract}

\section{Introduction}
\label{sec:intro}

Ideas from quantum information theory continue to play an important
role in the design and realisation of quantum computers and their
associated algorithms (see e.g.\ ref.~\cite{Nielsen:2012yss} for a
detailed review). There is also a growing community of people who are
interested in applying similar ideas to high-energy quantum systems
such as particle colliders. A key example of the latter is purported
tests of entanglement, with a canonical example being the production
of pairs of top quarks at the
LHC~\cite{Afik:2020onf,Dong:2023xiw,Aoude:2022imd,Fabbrichesi:2021npl,Severi:2021cnj,Afik:2022kwm,Aguilar-Saavedra:2022uye,Fabbrichesi:2022ovb,Afik:2022dgh,Severi:2022qjy,Aguilar-Saavedra:2023hss,Han:2023fci,Simpson:2024hbr,Aguilar-Saavedra:2024hwd,Maltoni:2024csn,ATLAS:2023jzs,ATLAS:2023fsd,CMS:2024hgo}
(see ref.~\cite{Barr:2024djo} for a more general review, and
refs.~\cite{Abel:1992kz,Abel:2025skj,Bechtle:2025ugc,Low:2025aqq} for alternative perspectives). More recently, other quantum information ideas
have been investigated in a collider 
setting~\cite{White:2024nuc,Liu:2025frx,Liu:2025qfl,Fabbrichesi:2025ywl,CMS:2025cim,Afik:2025ejh,Aoude:2025jzc}
(see also
refs.~\cite{Barr:2022wyq,AshbyPickering:2022umy,Aguilar-Saavedra:2022wam,Aguilar-Saavedra:2022mpg,Fabbri:2023ncz,Aoude:2023hxv,Fabbrichesi:2023cev,Sakurai:2023nsc,Altomonte:2023mug,Afik:2024uif,Aguilar-Saavedra:2024vpd,Aguilar-Saavedra:2024whi,Grabarczyk:2024wnk,Morales:2024jhj,Altomonte:2024upf,Han:2024ugl,Cheng:2024rxi,Subba:2024aut,Subba:2024mnl,Horodecki:2025tpn,DelGratta:2025qyp,Nason:2025hix,Grossi:2024jae,Cheng:2025cuv}
for related works), and in the theory of particle scattering and decay processes~\cite{Balasubramanian:2011wt, Seki:2014cgq, Peschanski:2016hgk,Grignani:2016igg,Kharzeev:2017qzs,Fan:2017hcd,Fan:2017mth,Cervera-Lierta:2017tdt,Beane:2018oxh,Rigobello:2021fxw,Low:2021ufv,Liu:2022grf,Fedida:2022izl,Cheung:2023hkq,Carena:2023vjc,Aoude:2024xpx,Low:2024mrk,Low:2024hvn,Thaler:2024anb,McGinnis:2025brt,Carena:2025wyh,Liu:2025pny,Hu:2025lua,Sou:2025tyf}, highlighting the various reasons for this increasing
interest. Quantum information methods may inform ways to distinguish
new physics from the Standard Model of Particle Physics, a point
emphasised particularly by
refs.~\cite{Aoude:2022imd,Aoude:2023hxv,Aoude:2025jzc,Fabbrichesi:2025ywl}. Going
the other way, physical systems studied by high-energy physicists
(however arcane or theoretical) may provide useful insights into open
questions that concern the more traditional quantum theory
community. It is this latter spirit that motivates the present study.

We will study a particular quantum information theory property called
{\it magic}, also known as {\it non-stabiliserness}. To understand the
latter term, it is sufficient to note that there is a well-defined
family of special quantum states --- called {\it stabiliser states} ---
that can be created using a particular type of quantum gate
(i.e.\ {\it Clifford gates}). It is known (see
e.g.\ ref.~\cite{Nielsen:2012yss}) that quantum algorithms containing
only stabiliser states lead to no computational advantage over
equivalent classical computer algorithms. Thus, to build a genuinely
powerful quantum computer, one must ensure that at least some
intermediate states are non-stabiliser. ``Magic'' quantifies the
degree to which this occurs, and is also known to be important for
building fault-tolerant algorithms, such that the study of magic is
related to the biggest open problems in real-world quantum
computing. As such, there are now even entire conferences devoted to
the study of magic in many-body systems.\footnote{The Many Body Quantum
Magic (MBQM) series originated in 2024, in Abu Dhabi, and has been held
in Seattle in 2025.}

Magic has been studied in a range of quantum
systems~\cite{Beverland_2020,PhysRevApplied.19.034052,Leone:2023avk,Qassim2021improvedupperbounds,Leone:2021rzd,Haug:2023ffp,Magic1,PhysRevA.108.042408,Gu:2023qqq,Tirrito:2023fnw,Turkeshi:2023lqu,Leone:2024lfr,Tarabunga:2023ggd,Frau:2024qmf,Lami:2024osd,Robin:2024bdz,Chernyshev:2024pqy},
and in ref.~\cite{White:2024nuc} was shown to be present in the same
top quark pair system at the LHC that had been previously used to
potentially probe entanglement and related new physics
signatures. Since then, ref.~\cite{Liu:2025frx} considered theoretical
upper bounds on magic production in two-qubit systems, and
ref.~\cite{Liu:2025qfl} looked at how efficiently one can generate
magic in $2\rightarrow 2$ scattering processes in
QED. refs.~\cite{Fabbrichesi:2025ywl,Aoude:2025jzc,Busoni:2025dns} have considered how information measures including magic can be used as probes of new physics, and there has also been experimental
verification that magic top quarks are routinely produced in
nature~\cite{CMS:2025cim}. Given the importance of magic for quantum
computing, and the mystery surrounding how to produce and enhance this
quantity in full generality for arbitrary quantum systems, there is a
clear role for case studies that look at how magic is produced in
given contexts, as has been done in refs.~\cite{White:2024nuc} (for
top quarks) and~\cite{Liu:2025qfl} (for QED processes). Furthermore,
it is interesting to try to look for closely related systems, and to
compare the profile of magic production when one or more physical
parameters is varied. This may in turn lead to important insights that
can be ported to other systems, including those in condensed matter
and / or optics.

The aim of this paper is to perform such a case study, focusing on the
particular context of $2\rightarrow 2$ scattering processes involving
gluons or gravitons. The relevant theories in each case are
Yang--Mills theory,\footnote{With gauge group SU(3), Yang--Mills theory
can be thought of as the theory of the strong force (Quantum
Chromodynamics) without the quarks. However, we will not restrict to a
particular gauge group unless otherwise stated.} and gravity in the
form of General Relativity. In both cases, the particles being
scattered have two polarisation states, and hence the initial and
final states form a two-qubit system. The effect of the Yang--Mills or
gravitational interactions is then to change the initial quantum
state, which can possibly lead to the creation of magic in spin
space. Furthermore, these theories provide an example of closely
related situations: in moving from Yang--Mills to gravity, the (zero)
mass and number of polarisation states of the scattering particles
remain the same, and only the spin changes: from spin one for gluons,
to spin two for gravitons. Thus, these two theories provide a good
example of two closely related types of theory, whose lessons may be
fruitful for other systems which are related by a change of spin. We
will see that both gluon and graviton scattering are capable of
generating magic states, even when no magic is present in the initial
state. The amount of magic depends upon the scattering angle and,
unsurprisingly, the profile of magic changes according to the nature
of the scattering particles. 

This leads us to another good reason to study Yang--Mills theory and
gravity, which is that their theoretical structures are known to be
mathematically very closely related. We will see this explicitly in
what follows, in which we will need scattering amplitudes relating
initial and final two-qubit states in the two theories. We will
restrict ourselves to tree-level scattering, and it is then known that
the relevant amplitudes in gravity can be obtained by those in
Yang--Mills theory by the so-called {\it Kawai--Lewellen--Tye relations}, originally
derived in string theory~\cite{Kawai:1985xq}, which state that gravity
amplitudes can be written as certain sums of products of gauge theory
ones. More recently, such relationships have been generalised to
loop-level scattering
amplitudes~\cite{Bern:2008qj,Bern:2010ue,Bern:2017yxu,Bern:2010yg} and
classical
solutions~\cite{Monteiro:2014cda,Luna:2018dpt,White:2020sfn,Chacon:2021wbr,Goldberger:2016iau,Anastasiou:2018rdx},
and the general programme for relating quantities in gauge and gravity
theories is known as the {\it double copy} (see
e.g.\ refs.~\cite{Bern:2019prr,Adamo:2022dcm,White:2024pve} for
reviews). This tight relationship between gauge and gravitational
physics means that we will be able to interpret {\it why} the profile
of magic generation differs between our chosen types of theory, and
why certain features are similar. This is as useful for interpreting
the physics of the double copy, as it is for drawing insights about
magic. Also, the fact that double-copy-like relationships have started
to be extended to other theories of possible interest in condensed
matter~\cite{White:2024pve}, suggests that our insights and
methods from this paper may be highly portable. To illustrate this further, we will supplement our analysis of Yang--Mills and gravity with results from supersymmetric generalisations of these theories, whose amplitudes can also be obtained using KLT relations. The additional particles -- gluinos and gravitinos -- have half-integer spin, and will be seen to fit the same pattern of decreasing magic with increasing spin.

The structure of our paper is as follows. In section~\ref{sec:review},
we review key concepts needed for what follows. In
section~\ref{sec:amps}, we show how to calculate magic in both
Yang--Mills theory and gravity, using known results from the study of
scattering amplitudes. We present
results for magic and other observables, comparing similarities and differences between the two theories. In section~\ref{sec:SUSY}, we broaden our results to particles of different spin, by using supersymmetric generalisations of Yang--Mills theory and gravity.
We discuss our results and conclude in section~\ref{sec:conclude}. 

\section{Review of necessary concepts}
\label{sec:review}

In this section, we review salient details regarding magic of quantum
states. For a fuller review of magic in a collider context, we refer
the reader to ref.~\cite{White:2024nuc}.  Our starting point is to
consider an $n$-qubit system, where each qubit has basis states
$|0\rangle$ and $|1\rangle$, such that a basis for the full $n$-qubit
Hilbert space is provided by the states
\begin{equation}
  |i\rangle\otimes|j\rangle\otimes\ldots\otimes|l\rangle\,.
  \label{statecomp}
\end{equation}
One may then consider the family of {\it Pauli string} operators
\begin{equation}
  {\cal P}_n=P_1\otimes P_2\otimes \ldots\otimes P_N,\quad
  P_a\in\{\mathbbm{1},\sigma_1,\sigma_2,\sigma_3\}\,,
  \label{Paulistring}
\end{equation}
such that a Pauli or $2\times 2$ identity matrix acts on each
individual qubit. The Pauli strings then generate the \emph{Pauli group}, which consists of the Pauli strings weighted by phases of $\pm 1,\pm i$. For $n$ qubits, we define the \emph{Clifford group} $\mathcal{C}_n$ as the normaliser of the Pauli group in \(\text{U}(2^n)\); concretely,
\begin{equation}
    \mathcal{C}_n \equiv \{U \in \text{U}(2^n):U P U^\dagger = e^{i\theta} P'\}\,,
\end{equation}
where $P$ and $P'$ are Pauli strings and $\theta\in\{0,\pi/2,\pi,3\pi/2\}$. In quantum computing and quantum information theory, the members of this group are referred to as \emph{Clifford gates}.

A general quantum state $|\psi\rangle$ can be
classified by its {\it Pauli spectrum}
\begin{equation}
  {\rm spec}(|\psi\rangle)=\{\langle\psi|P|\psi\rangle,\quad P\in
  {\cal P}_n\}\,,
  \label{Paulispec}
\end{equation}
namely by the set of $4^n$ expectation values of each Pauli
string. For a given $n$, a special set of {\it stabiliser states} can
be defined, whose Pauli spectrum has $2^n$ values equal to $\pm1$, and
the rest zero. Physically, these are the states obtained by acting on
the state
\begin{displaymath}
  |0\rangle\otimes|0\rangle\otimes\ldots\otimes|0\rangle
\end{displaymath}
with Clifford gates. An
important result known as the Gottesmann--Knill theorem states that
a quantum circuit comprised of only Clifford circuits (thus generating only stabiliser states) may be simulated in polynomial time on a classical computer. It is therefore useful to
have a property that measures the ``non-stabiliserness'' of a quantum
state $|\psi\rangle$, which can in turn be used to measure the degree
of quantum computational advantage a given quantum circuit has over a
classical computer. This property has become known as {\it magic} in
the quantum information literature and can be defined in terms of the
Pauli spectrum of $|\psi\rangle$, given that this is what decides
whether a given state is stabiliser or not. Various definitions of
magic exist, and we will here adopt the family of
{\it Stabiliser R\'{e}nyi Entropies} (SRE) of ref.~\cite{Leone:2021rzd}. For
the general case of mixed quantum states, these can be written as
\begin{equation}
  M_q(|\psi\rangle)=-\frac{1}{1-q}\log_2\left(\frac{\sum_{P\in{\cal P}_n} \langle \psi|
  P|\psi\rangle^{2q} }
  {\sum_{P\in{\cal P}_n}\langle\psi|P|\psi\rangle^2}\right)\,.
  \label{Mqdef}
\end{equation}
There is a different SRE for each integer value $q\geq 2$, and the
numerator and denominator in the logarithm can be seen to explicitly
contain the Pauli spectrum, where there is a sum over all the
different possible Pauli strings. One may show that the quantities of
eq.~(\ref{Mqdef}) indeed vanish for stabiliser states, and are also
additive when combining quantum systems. One may think of the set of
values $\{M_q\}$ as characterising the moments of the Pauli spectrum
of $|\psi\rangle$, and it is often sufficient to use the Second
Stabiliser R\'{e}nyi entropy (SSRE) to quantify non-zero magic, as has
been done in a collider context in
refs.~\cite{White:2024nuc,Aoude:2025jzc,Fabbrichesi:2025ywl}.

In this work, we wish to analyse the \(2\to 2\) scattering of gluons. At leading order in perturbation theory, states of more than two gluons are not accessible, and the full multi-particle Fock space reduces to a two-particle Hilbert space. Generalising the approach of ref.~\cite{Kowalska:2024kbs}, we partition this Hilbert space as \(\mathcal{H}\simeq\mathcal{H}_\text{hel}\otimes\mathcal{H}_\text{col}\otimes\mathcal{H}_\text{kin}\), where \(\mathcal{H}_\text{hel}\simeq \mathbb{C}^2\otimes\mathbb{C}^2\) corresponds to the discrete helicity degrees of freedom, \(\mathcal{H}_\text{col}\simeq\mathbb{C}^d\otimes\mathbb{C}^d\) to the discrete adjoint degrees of freedom of the $d$-dimensional gauge group, and \(\mathcal{H}_\text{kin}\simeq L^2(\mathbb{R}^3\otimes\mathbb{R}^3)\) to the continuous momentum degrees of freedom. States in these spaces are normalised in the canonical way:
\begin{equation}
    \braket{J|K}=\delta_{JK},\quad \braket{ab|cd}=\delta_{ac}\delta_{bd},\quad\braket{\mathbf{p}_1\mathbf{p}_2|\mathbf{p}_3\mathbf{p}_4}=4E_1 E_2(2\pi)^6\delta^{(3)}(\mathbf{p}_3-\mathbf{p}_1)\delta^{(3)}(\mathbf{p}_4-\mathbf{p}_2)\,,
\end{equation}
where \(\{\ket{J}\}\equiv\{\ket{00},\ket{01},\ket{10},\ket{11}\}\) is a basis for the helicity space, and $a$, $b$ etc. denote adjoint (colour) indices. The completeness relation is then fixed to
\begin{equation}
    \mathbbm{1}=\sum_{J}\sum_{ab}\int\frac{d^3\mathbf{p}_1}{(2\pi)^3}\frac{d^3\mathbf{p}_2}{(2\pi)^3}\frac{1}{4E_1 E_2}\ket{J;ab;\mathbf{p}_1\mathbf{p}_2}\!\bra{J;ab;\mathbf{p}_1\mathbf{p}_2}\,.
\end{equation}
We assume an initial state of the form
\begin{equation}
    \label{eq:in-state}
    \ket{\text{in}}=\frac{1}{\sqrt{V}}\ket{J}\otimes\ket{a_1 a_2}\otimes\ket{\mathbf{p}_1\mathbf{p}_2}\equiv\frac{1}{\sqrt{V}}\ket{J;a_1 a_2;\mathbf{p}_1\mathbf{p}_2}\,,
\end{equation}
where \(V=4E_1 E_2(2\pi)^6\delta^{(3)}(0)\delta^{(3)}(0)\) is a normalisation such that \(\braket{\text{in}|\text{in}}=1\).\footnote{One could imagine using states that are not perfectly sharp in momentum; then the normalisation would not be formally infinite. In any event, this normalisation will factor out shortly.} The final state is then given by \(\ket{\text{out}}=S\ket{\text{in}}\), where \(S=\mathbbm{1}+iT\) is the \(S\)-matrix. We project out a particular choice of final-state colour indices \(a_3,a_4\) and momenta \(\mathbf{p}_3,\mathbf{p}_4\); this corresponds to a fictitious \emph{measurement} of these quantum numbers.\footnote{Gluons (and quarks) are confined, and thus the asymptotic Hilbert space consists of colour singlets. At high energies, though—before hadronisation takes over—it is mathematically convenient to work with states that carry definite colour indices. These states should be understood as formal artifacts: they are not physical asymptotic states, but they allow us to compute partonic cross–sections and, in the present context, quantum–information–theoretic quantities.} Assuming \(a_3 a_4\neq a_1 a_2\), it can be shown using the completeness relation that the projected state is
\begin{align}
    \label{eq:proj}
    \ket{\psi}&=\ket{a_3 a_4;\mathbf{p}_3\mathbf{p}_4}\!\braket{a_3 a_4;\mathbf{p}_3\mathbf{p}_4|\text{out}} \nonumber \\[6pt]
    &=\frac{(2\pi)^4\delta^{(4)}(p_1+p_2-p_3-p_4)}{\sqrt{V}}\sum_{K}i\mathcal{A}(J\to K)\ket{K;a_3 a_4;\mathbf{p}_3\mathbf{p}_4}\,,
\end{align}
introducing the shorthand notation for the scattering amplitude
\begin{equation}
    \label{a'Kdef2}
    (2\pi)^4\delta^{(4)}(p_1+p_2-p_3-p_4)i\mathcal{A}(J\to K)\equiv\braket{K;a_3 a_4;\mathbf{p}_3\mathbf{p}_4|iT|J;a_1 a_2;\mathbf{p}_1\mathbf{p}_2}
\end{equation}
(again, fixing \(a_1,a_2,a_3,a_4\) and the external momenta). Since we have performed this projection, we are required to explicitly normalise the state; we thus write
\begin{equation}
    \label{eq:norm}
    \ket{\psi}\to\frac{1}{\sqrt{\braket{\psi|\psi}}}\ket{\psi}=\frac{1}{\sqrt{\braket{\psi|\psi}}}\sum_{K}i\mathcal{A}(J\to K)\ket{K;a_3 a_4;\mathbf{p}_3\mathbf{p}_4}\,.
\end{equation}
Going forward, we will suppress the colour and momentum labels when writing the states, but their presence should be implicitly assumed. We note that the combination of (i) starting in an initial state of the form of eq.~\eqref{eq:in-state}, (ii) the projection of eq.~\eqref{eq:proj} onto a colour computational basis state, and (iii) the necessary normalisation step of eq.~\eqref{eq:norm} are sufficient ingredients to ensure that the out state $| \psi \rangle$ has no colour dependence. That is, non-vanishing out states subject to the aforementioned criteria are equal (up to a phase) for different initial colour states and post-scattering projections. This feature was called \textit{universality} in ref.~\cite{Nunez:2025dch}. We have verified explicitly that universality is {\it not} realised if the initial colour state is a superposition of states with different colour indices.

\section{Magic in Yang--Mills theory and gravity}
\label{sec:amps}

\begin{figure}
  \begin{center}
    \scalebox{0.5}{\includegraphics{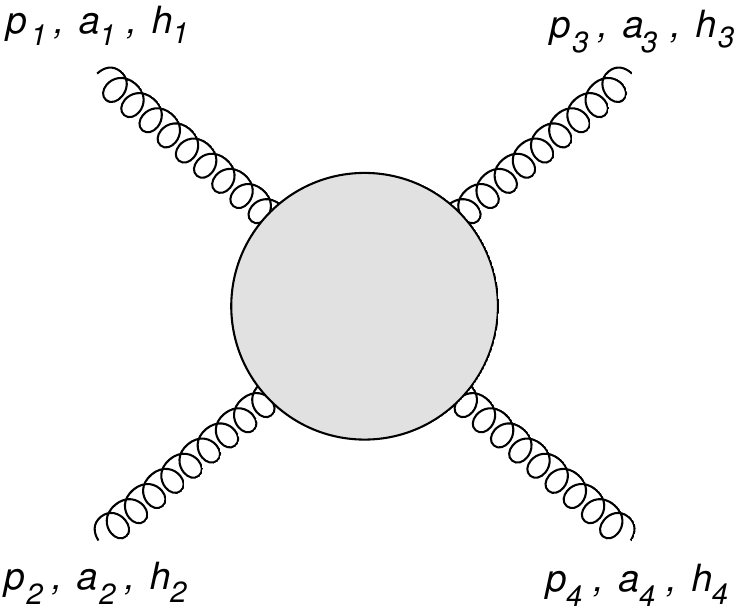}}
    \caption{The $2\rightarrow 2$ scattering of gluons, where
      $\{p_i\}$, $\{a_i\}$ and $\{h_i\}$ label momenta, colour indices
      and helicities respectively. Initial momenta are taken to be
      incoming, and final state momenta outgoing.}
    \label{fig:gluons}
  \end{center}
\end{figure}

Having introduced the framework for calculating magic in $2\rightarrow
2$ scattering processes of qubits, let us now turn to the particular
case of gluon scattering. More precisely, we will consider the process
shown in figure~\ref{fig:gluons}, such that we must keep track of the
momentum $p_i$ of each gluon and its colour index $a_i$. For both the
initial and final states, we must also choose a map from spin space to
physical space, thus defining the meaning of the states $|0\rangle$
and $|1\rangle$ for each qubit. For both the initial and final states,
we will take $|0\rangle$ and $|1\rangle$ to be the gluon states of
positive and negative helicity respectively, where the helicity
measures whether the spin is (anti-)aligned with the momentum
direction. In physical terms, the positive (negative) helicity state
corresponds to right (left) circular polarisation of the gluon. From
now on, we will make our choice of basis clear by relabelling as follows:
\begin{equation}
  |J\rangle\in\{|++\rangle,|+-\rangle,|-+\rangle,|--\rangle\}\,.
  \label{Jbasis}
\end{equation}
We further denote the helicity of gluon $i$ when needed by $h_i$, as
shown in figure~\ref{fig:gluons}. We take the initial state momenta to
be {\it incoming}, and the final state momenta to be {\it outgoing},
such that $p_i$ corresponds to the 4-momentum one would observe in the
physical scattering process. Our reason for stressing this is that
relevant results in the QFT literature are often written down for all
outgoing momenta. To convert from an outgoing initial state gluon to
an incoming one, one must take $p_i\rightarrow -p_i$ and
$h_i\rightarrow -h_i$ (i.e.\ the helicity also flips).

In our chosen basis, the scattering amplitudes appearing in
eq.~(\ref{a'Kdef2}) will be so-called {\it helicity amplitudes}, in
which the helicity of all particles is fixed. That this choice is
particularly convenient stems from the fact that helicity amplitudes
are very well-studied quantities in both Yang--Mills theory and gravity
(see e.g.\ ref.~\cite{Elvang:2015rqa,Henn:2014yza,Schwartz:2014sze} for
excellent contemporary reviews of helicity methods for scattering
amplitudes). A first useful result is that the total tree-level result
for an $n$-point amplitude (with all outgoing momenta) can be written
as
\begin{equation}
  {\cal A}^{\rm tot.}_n[1^{h_1}2^{h_2}\ldots n^{h_n}]
  =g^{n-2}\sum_{{\rm perms} \ \sigma}
  {\cal A}_n[1^{h_1}\sigma(2^{h_2}\ldots n^{h_n})]{\rm Tr}
  \left[{\bf T}^{a_1}\sigma({\bf T}^{a_2}\ldots{\bf T}^{a_n})\right]\,.
  \label{Atot}
\end{equation}
We have here used a common notation in which e.g.\ ${\cal
  A}_4[1^+2^+3^+ 4^+]$ represents a 4-point amplitude with all
positive helicities, and colour indices and momenta labelled according
to figure~\ref{fig:gluons} (but with all momenta outgoing). On the
right-hand side of eq.~(\ref{Atot}), $g$ is the coupling constant,
${\bf T}^{a_1}$ a colour generator, and the sum is over all
permutations $\sigma$ of $(2\ldots n)$. The quantity
\begin{displaymath}
{\cal A}_n[1^{h_1}\sigma(2^{h_2}\ldots n^{h_n})]
\end{displaymath}
is called a {\it colour-ordered amplitude}, due to the fact that it is
the coefficient of a particular ordering of colour generators. Much is
known about these quantities. For example, one may show that they
vanish if either: (i) all helicities are equal (for outgoing momenta);
(ii) all but one of the helicities are equal (for outgoing momenta). Thus,
the first non-zero amplitudes occur if two helicities are different to
all the rest, and such amplitudes are thus called {\it maximally
  helicity violating (MHV)}. For our special case of $2\rightarrow 2$
scattering, we have $n=4$, and the only non-vanishing colour-ordered
amplitudes will have (for all outgoing momenta) two positive helicity
states, and two negative ones. Translating back to incoming momenta in
the initial state, this means that the only non-zero amplitudes are
\begin{align*}
  &{\cal A}({++\rightarrow ++})\,,\quad
  {\cal A}({--\rightarrow --})\,,\quad
  {\cal A}({+-\rightarrow +-})\,,\notag\\
  &\quad
  {\cal A}({+-\rightarrow -+})\,,\quad
  {\cal A}({-+\rightarrow -+})\,,\quad
  {\cal A}({-+\rightarrow +-})\,. 
\end{align*}
A known result exists for tree-level MHV amplitudes, for any number of
gluons. Returning to the case of all-outgoing momenta, one has the
so-called {\it Parke-Taylor formula}~\cite{Parke:1986gb}
\begin{equation}
  {\cal A}_n[1^+\ldots i^-\ldots j^-\ldots n^+]=\frac{\langle ij\rangle^4}
  {\langle 12\rangle \langle 23\rangle\ldots \langle n1\rangle}\,,
  \label{Parke-Taylor}
\end{equation}
written in terms of {\it spinor products}
\begin{equation}
  \langle ij\rangle=e^{i\phi_{ij}}\sqrt{(p_i+p_j)^2}\,,
  \label{spinprod}
\end{equation}
defined for outgoing momenta, and where $\phi_{ij}$ is a real
phase. This provides all the ingredients needed for computing the
amplitude matrix appearing in eq.~(\ref{a'Kdef2}). First, one may
evaluate all Parke-Taylor amplitudes using eq.~(\ref{spinprod}) which,
upon switching back to incoming momenta in the initial state can be
written in terms of the Mandelstam invariants
\begin{equation}
  s=(p_1+p_2)^2\,,\quad t=(p_1-p_3)^2\,,\quad u=(p_1-p_4)^2\,.
  \label{mandies}
\end{equation}
It is also convenient to transform from the colour basis appearing in
eq.~(\ref{Atot}) to one involving explicit products of structure
constants, and a particularly convenient choice is the Del-Duca--Dixon--Maltoni (DDM) basis \cite{DelDuca:1999rs}, given in terms of two independent colour structures
\begin{equation}
  c_{1234}=f^{a_1 a_2 c}f^{a_3 a_4 c}\,,\quad c_{1324}=f^{a_1 a_3 c}f^{a_2 a_4 c}\,.
  \label{F12def}
\end{equation}
In this basis, the four-gluon amplitude is given by
\[
\cl{A}[1^{h_1}2^{h_2}3^{h_3}4^{h_4}] = \left[c_{1234}\mathcal{A}_4[1^{h_1}2^{h_2}3^{h_3}4^{h_4}] + c_{1324}\mathcal{A}_4[1^{h_1}3^{h_3}2^{h_2}4^{h_4}]\right]\,,
\]
where colour-ordered amplitudes $\mathcal{A}_4$ can be computed via the Parke-Taylor formula in eq.~\eqref{Parke-Taylor}. Using the relation between spinor products and Mandelstam variables, we can express the amplitudes as
\begin{align}
  {\cal A}({++\rightarrow ++})=  {\cal A}({--\rightarrow --})
  &=g^2\left[
    c_{1234}\left(\frac{s}{u}\right)
    +c_{1324}\left(\frac{s^2}{tu}\right)
    \right]\,;\notag\\
  {\cal A}({+-\rightarrow +-})=  {\cal A}({-+\rightarrow -+})
  &=g^2\left[
    c_{1234}\left(\frac{u}{s}\right)
    +c_{1324}\left(\frac{u}{t}\right)
    \right]\,;\notag\\
  {\cal A}({+-\rightarrow -+})=  {\cal A}({-+\rightarrow +-})
  &=g^2\left[
    c_{1234}\left(\frac{t^2}{su}\right)
    +c_{1324}\left(\frac{t}{u}\right)
    \right]\,.
  \label{Aresults}
\end{align}
where we have restricted ourselves to real, $2\rightarrow 2$ kinematics where the phase $\phi_{ij}$ in eq.~\eqref{spinprod} is given by 
\[
\phi_{ij} = (1-\Theta(\braket{ij}))\pi\,,
\]
with $\Theta(x)$ the Heaviside function.

\begin{figure}[t]
  \begin{center}
    \includegraphics[width=0.70\linewidth]{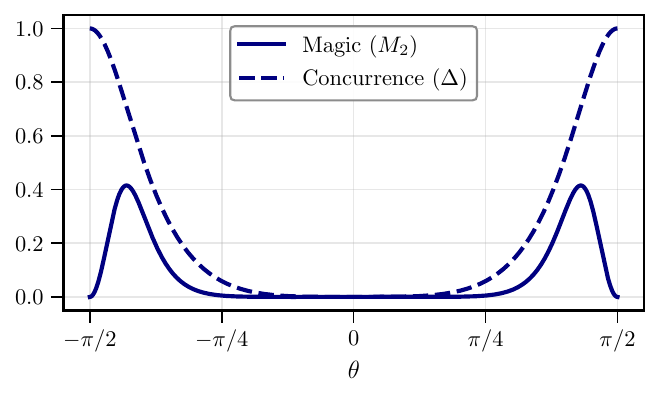}
    \caption{The magic $M_2$ (solid) and concurrence (dashed) of the
      final state obtained from the initial gluon state $|+-\rangle$,
      as a function of the scattering angle $\theta$.}
    \label{fig:M2Delta}
  \end{center}
\end{figure}

Scattering amplitudes in quantum General Relativity may be defined by
expanding the full metric of spacetime in terms of the (vacuum)
Minkowski part, plus a correction:
\begin{equation}
  g_{\mu\nu}=\eta_{\mu\nu}+\kappa h_{\mu\nu}\,.
  \label{gmunu}
\end{equation}
Here $\kappa=\sqrt{32\pi G_N}$ in terms of the Newton constant $G_N$,
and $h_{\mu\nu}$ is known as the graviton field, representing a spin
two and massless particle. By expanding the action for General
Relativity in terms of the graviton, one obtains an infinite series of
graviton interactions, and can apply the language of quantum field
theory and Feynman rules in order to calculate scattering
processes.\footnote{General Relativity is non-renormalisable, but can
be treated as a highly convergent effective theory involving
increasingly complicated graviton interactions order-by-order in
perturbation theory.} It turns out, however, that there is a
remarkably simple way to obtain $2\rightarrow 2$ graviton amplitudes
directly from the corresponding Yang--Mills ones. First, one may note
that each graviton has two helicity states, so that one may define the
basis of eq.~(\ref{Jbasis}) also for gravitons. It then turns out that
the same rules apply regarding individual helicity amplitudes, such
that only MHV configurations are non-zero at four points. Finally, the
tree-level four-point graviton amplitude for any external helicity
configuration is given by a product of colour-ordered amplitudes
\begin{equation}
  \mathcal{A}^{(2)}[1^{h_1}2^{h_2}3^{h_3}4^{h_4}]=
  -\frac{\kappa}{2} t \mathcal{A}_4^{(1)}[1^{h_1}2^{h_2}3^{h_3}4^{h_4}]
  \mathcal{A}_4^{(1)}[1^{h_1}3^{h_3}2^{h_2}4^{h_4}]\,.
  \label{M4def}
\end{equation}
(We hereafter adopt the notation \(\mathcal{A}^{(s)}\) for the amplitudes, where \(s\) denotes the spin of the appropriate particle species.) This is a special case of a general family of relations known as the
{\it KLT relations}~\cite{Kawai:1985xq}, which relate graviton
amplitudes to sums of products of colour-ordered gluon
amplitudes. Note, however, that colour degrees of freedom themselves
are absent in eq.~(\ref{M4def}), as they should be given that there is
no colour in gravity. Furthermore, the Yang--Mills coupling has been
replaced with the gravitational coupling (up to a numerical factor). The KLT relations originated in string
theory, in which closed string amplitudes can be written as products
of open string amplitudes. Upon taking the low energy limit, closed
strings give rise to gravitons (roughly speaking), and open strings
give rise to gluons, leading to results such as eq.~(\ref{M4def}). As
mentioned in the introduction, more recent research has shown that
relations between gauge theory and gravity go much
further~\cite{Bern:2008qj,Bern:2010ue,Bern:2017yxu,Bern:2010yg}, such
that studies of quantum information in the different theories can also
be generalised significantly beyond the initial analysis of this
paper. Applying eq.~(\ref{M4def}) to the present case of tree-level
$2\rightarrow 2$ scattering, we find non-zero graviton amplitudes (for
incoming initial state momenta)
\begin{align}
  \mathcal{A}^{(2)}({++\rightarrow ++})=  \mathcal{A}^{(2)}({--\rightarrow --})
  &=-\frac{\kappa^2}{2} \left(\frac{s^3}{ut}\right)\,;\notag\\
  \mathcal{A}^{(2)}({+-\rightarrow +-})=  \mathcal{A}^{(2)}({-+\rightarrow -+})
  &=-\frac{\kappa^2}{2} \left(\frac{u^3}{st}\right)\,;\notag\\
  \mathcal{A}^{(2)}({+-\rightarrow -+})=  \mathcal{A}^{(2)}({-+\rightarrow +-})
  &=-\frac{\kappa^2}{2} \left(\frac{t^3}{su}\right)\,.
  \label{Mresults}
\end{align}
Eqs.~(\ref{Aresults}, \ref{Mresults}) may be used in
eq.~(\ref{a'Kdef2}) to convert an arbitrary initial gluon or graviton state
into an unnormalised final state, from which one may calculate the
magic according to eq.~(\ref{Mqdef}) after normalisation.
In presenting results, we will begin by taking a
particular initial state $|+-\rangle$. This has been considered
recently in ref.~\cite{Liu:2025qfl}, in the context of studying how
entanglement can be generated in gluon scattering. The $|++\rangle$
and $|--\rangle$ initial states create no entanglement in the final
state, and results from $|-+\rangle$ are the same as those from
$|+-\rangle$, due to the relations between amplitudes in
eqs.~(\ref{Aresults}, \ref{Mresults}). Given that it is interesting to
compare magic and entanglement in the final state, let us quantify the
latter, as in ref.~\cite{Liu:2025qfl}, using the {\it concurrence},
which in our present notation reads
\begin{equation}
  \Delta=|a_{++}a_{--}-a_{+-}a_{-+}|\,,
  \label{Deltadef}
\end{equation}
where it is understood that the final state $\ket{\psi}\equiv\sum_{J}a_{J}\!\ket{J}$ has
already been normalised. We may then evaluate both the concurrence and
magic (as measured by the SSRE $M_2$) in the centre-of-mass frame, in
which the momenta are given by
\begin{align}
  p_1&=(E,0,0,E)\,;\notag\\
  p_2&=(E,0,0,-E)\,;\notag\\
  p_3&=(E,E\sin\theta,0,E\cos\theta)\,;\notag\\
  p_4&=(E,-E\sin\theta,0,-E\cos\theta)\,,
  \label{pparam}
\end{align}
where $E$ is the energy of each gluon or graviton, and $\theta$ the
scattering angle. The Mandelstam invariants are then given by
\begin{equation}
  s=4E^2\,,\quad t=-4E^2\sin^2\left(\frac{\theta}{2}\right)\,,\quad
  u=-4E^2\cos^2\left(\frac{\theta}{2}\right)\,.
  \label{mandies2}
\end{equation}
For our particular final state, we find that the concurrence and magic
are given by
\begin{equation}
  \Delta(|\psi\rangle)=\frac{2 t^2 u^2}{t^4+u^4}\,,\quad
  M_2(|\psi'\rangle)=-\log_2\left(
  \frac{t^{16}+14t^8u^8+u^{16}}{(t^4+u^4)^4}
  \right).
  \label{DeltaM2}
\end{equation}
The first equation agrees with the result found in
ref.~\cite{Liu:2025qfl}, including the universality in colour mentioned above.  A plot of the
concurrence and magic is shown in fig.~\ref{fig:M2Delta}, as a
function of the scattering angle. The entanglement rises from minimal
in the forward direction $\theta=0$, to maximal in the central
direction $\theta=\pi/2$, as noted in ref.~\cite{Liu:2025qfl}. The
magic profile, however, is very different. It rises slowly from zero
at $\theta=0$, reaching a maximum before decaying to zero at
$\theta=\pi/2$. Indeed, this can be understood by examining the explicit form of the final state at extremal angles:
\begin{equation}
  | \psi(\theta=0)\rangle=|+-\rangle\,,\quad
  | \psi(\theta=\pi/2)\rangle=\frac{1}{\sqrt{2}}\Big(|+-\rangle
  +|-+\rangle
  \Big)\,.
  \label{extremestates}
\end{equation}
Both of these turn out to be stabiliser states, and hence the magic
vanishes. Indeed, this follows a pattern previously seen in top quark
pair production, whose final state consists of two massive spin-1/2
particles~\cite{White:2024nuc}. There, magic is large when
entanglement is low, and vanishes in extremal regions of scattering
angle. The detailed relationship between magic and entanglement is an
ongoing research area, and our results confirm the view that magic
provides very different information to entanglement in general. 

\begin{figure}[t]
        \centering
        \includegraphics[width=0.70\linewidth]{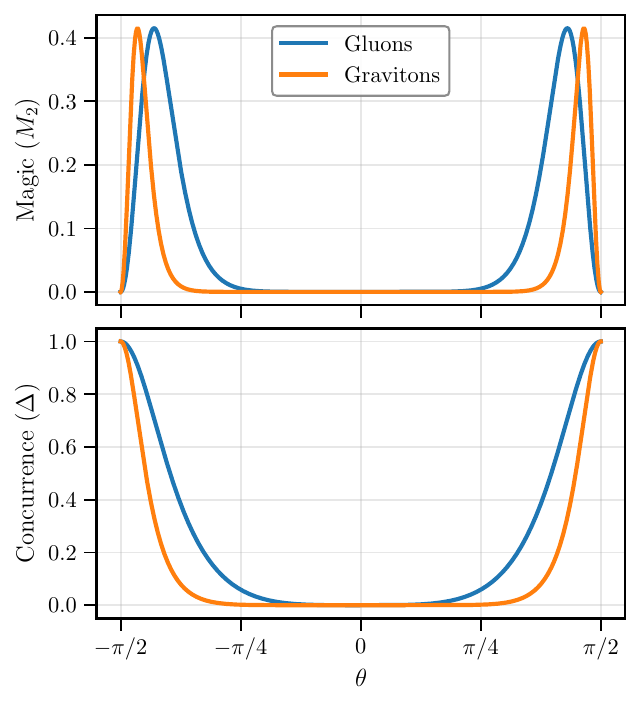}
        \caption
        {(Top) Magic $M_2$ of the final state obtained from initial
          state $|+-\rangle$ for gluons (blue) and gravitons (orange); (bottom) similar, but for the concurrence.}
        \label{fig:gluongrav1}
\end{figure}

In the top panel of figure~\ref{fig:gluongrav1}, we compare the magic for gluons and
gravitons, for the same final state as above (i.e.\ that obtained from
the $|+-\rangle$ initial state).
Interestingly, the qualitative shape of the magic profile is the same
in both theories, but the region of non-zero magic is much more
concentrated in gravity than in Yang--Mills theory. This can be traced
directly to the fact that gravity amplitudes can be obtained by {\it
  multiplying} Yang--Mills amplitudes, via the KLT relations. This
introduces higher powers of Mandelstam invariants, as can be seen from
the fact that the gravitational equivalent of eq.~(\ref{DeltaM2}) is
found to be
\begin{equation}
  M_2(|\psi\rangle_{\rm grav.})=
  -\log_2\left(
  \frac{t^{32}+14t^{16}u^{16}+u^{32}}{(t^8+u^8)^4}
  \right)\,.
  \label{M2grav}
\end{equation}
The higher powers in the numerator of the logarithm give rise to a
slower rise of the magic from $\theta=0$, as can be seen by comparing
the Taylor expansions of the magic from eq.~(\ref{DeltaM2},
\ref{M2grav}):
\begin{align}
  M_2(|\psi\rangle_{\rm YM})&=\frac{\theta^8}{64\log(2)}
  +{\cal O}(\theta^{10})\,;\notag\\
  M_2(|\psi\rangle_{\rm grav.})&=\frac{\theta^{16}}{16384\log(2)}
  +{\cal O}(\theta^{18})\,.
  \label{M2expand}
\end{align}
Given that double-copy like relationships are by no means limited to
gauge theory and gravity, it may well be the case that a similar
mechanism leads to concentration of magic for different types of qubit
in other physical systems. For this particular final state, it is also
interesting that the maximal value of the magic does not change in
moving from Yang--Mills to gravity, as can clearly be seen in the top panel of
fig.~\ref{fig:gluongrav1}. This will not turn out to be true for
less special initial states, as we will see below. Before moving on,
however, it is interesting to note that the concentration effect in
moving to gravity is also seen for entanglement (for our particular
initial state), as shown in the bottom panel of fig.~\ref{fig:gluongrav1}.

\begin{figure}[t]
  \begin{center}
    \includegraphics[width=0.7\linewidth]{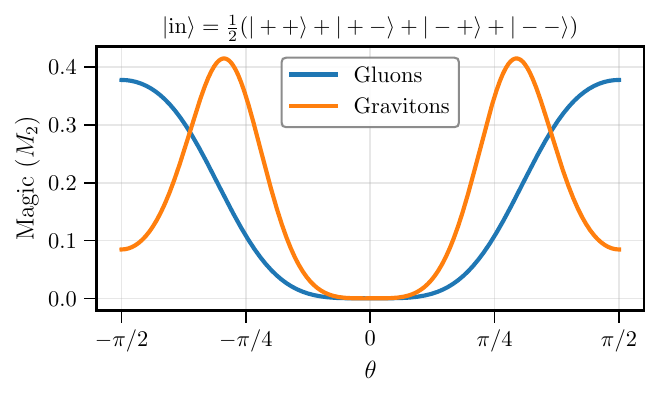}
    \caption{Magic $M_2$ of the final state arising from the initial
      stabiliser state of eq.~(\ref{stabstate1}), for gluons (blue)
      and gravitons (orange).}
    \label{fig:gluongrav2}
  \end{center}
\end{figure}

Inspired by ref.~\cite{Liu:2025qfl}, we can examine more general cases
of magic by taking for the initial state each of the 60 two-qubit
stabiliser states.\footnote{A full list of coefficients for the
stabiliser states ($\{a_J\}$ in our notation) can be found in appendix
A of ref.~\cite{Liu:2025qfl}.} Given that the magic of the initial
state must then be zero by definition, any magic in the final state
must be generated by the scattering process itself. It is not
difficult to find cases in which the gravity magic profile is
qualitatively different to the gluon one, and an example is shown in
fig.~\ref{fig:gluongrav2} based on the initial stabiliser state
\begin{equation}
  |\psi\rangle=\frac12\Big(
  |++\rangle+|+-\rangle+|-+\rangle+|--\rangle
  \Big)\,.
  \label{stabstate1}
\end{equation}
Physically, this corresponds to the case of unpolarised gluon or
graviton beams, and we see in this case that the magic is
non-vanishing in the central region $\theta=\pi/2$ for both gluons and
gravitons. However, the magic is maximised for gluons in the central
region, but has a local minimum for gravity.  

\begin{figure}[t]
  \begin{center}
    \includegraphics[width=0.70\linewidth]{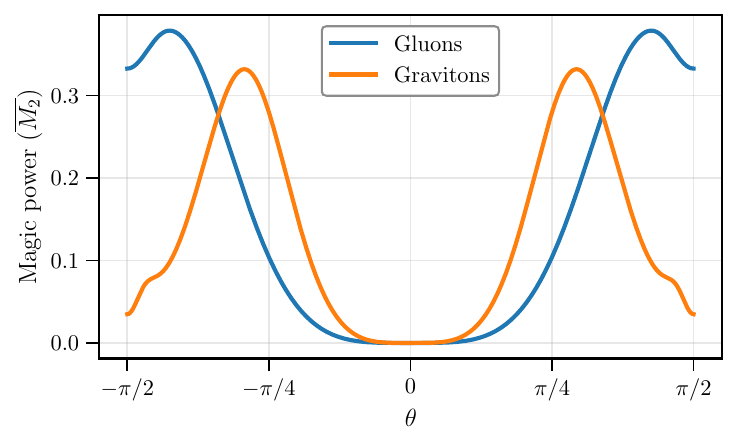}
    \caption{The magic power of eq.~(\ref{magicpow}) for gluons (blue)
      and gravitons (orange), as a function of scattering angle
      $\theta$.}
    \label{fig:magpow}
  \end{center}
\end{figure}

\begin{figure}[th!]
        \centering
        \begin{subfigure}[b]{0.49\textwidth}
            \centering
            \includegraphics[width=0.99\textwidth]{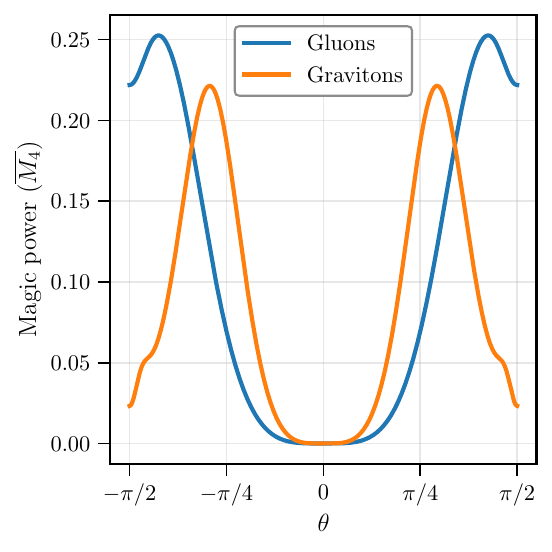}
            \caption{}%
        \end{subfigure}
        \hfill
        \begin{subfigure}[b]{0.49\textwidth}  
            \centering 
            \includegraphics[width=0.99\textwidth]{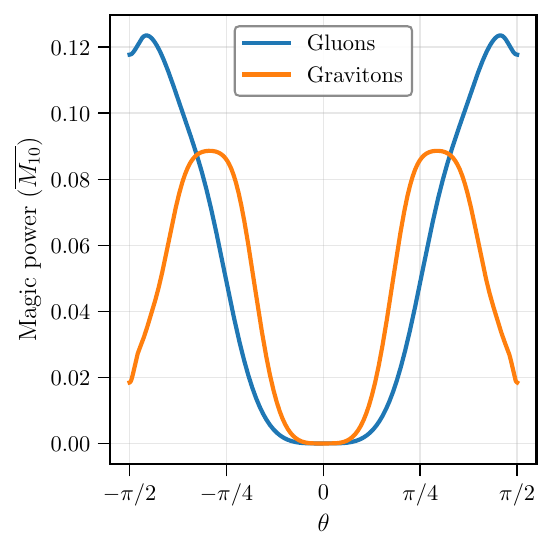}
            \caption{}%
        \end{subfigure}
        \caption
        {(a) The magic power $\overline{M_4}$ as a function of
          scattering angle, for gluons (blue) and gravitons (orange);
          (b) similar, but for $\overline{M_{10}}$.}
        \label{fig:M4M10}
\end{figure}

This leads us to examine the ``amount'' of magic in Yang--Mills theory
and gravity. One way to examine this is to look at the maximum value
of magic in the final state, upon cycling over all stabiliser states
in the initial state. Numerically, we find values
\begin{equation}
  M^{\rm max}_2\Big|_{\rm YM}=0.530\,,\quad
  M^{\rm max}_2\Big|_{\rm grav.}=0.415\,.
  \label{M2max}
\end{equation}
This suggests that the typical amount of magic is less in gravity than
in Yang--Mills theory, and indeed this can be made more precise by considering the
{\it magic power}, defined as the mean final state magic obtained from
all stabiliser initial states. Denoting the set of stabiliser states
by ${\cal S}$ and the final state corresponding to a given initial
state $\xi$ as $|\psi(\xi)\rangle$, we may write the magic power as
\begin{equation}
  \overline{M_2}=\frac{1}{60}\sum_{\xi\in{\cal S}}M_2(|\psi(\xi)\rangle),
  \label{magicpow}
\end{equation}
and find the following results in Yang--Mills theory and gravity:
\begin{align}
  \overline{M_2}\Big|_{\rm YM}&=
  -\frac{1}{15}\left[2\log_2\left(\frac{s^{16}+14s^8(t^2-u^2)^4+(t^2-u^2)^8}
    {(s^4+(t^2-u^2)^2)^4}\right)\right. \notag \\
    &\left.\quad+2\log_2\left(\frac{s^{16}+14s^8(t^2+u^2)^4+(t^2+u^2)^8}
    {(s^4+(t^2+u^2)^2)^4}\right)\right.\notag\\
    &\left.\quad+
    8\log\left(
    \frac{s^{16}+14s^8(t^8+u^8)+t^{16}+14t^8 u^8+u^{16}}
         {(s^4+t^4+u^4)^4}\right)\right.\notag\\
         &\left.\quad+\log_2\left(\frac{t^{16}+14t^8u^8+u^{16}}{(t^4+u^4)^4}\right)
         \right];
\end{align}
\begin{align}
  \overline{M_2}\Big|_{\rm grav.}&=
  -\frac{1}{15}\left[
    2\log_2\left(\frac{s^{32}+14s^{16}(t^4-u^4)^4+(t^4-u^4)^8}
    {(s^8+(t^4-u^4)^2)^4}\right)\right.\notag\\
    &\left.\quad+\log_2\left(\frac{s^{32}+14s^{16}(t^4+u^4)^4+(t^4+u^4)^8}
    {(s^8+(t^4+u^4)^2)^4}\right)\right.\notag\\
    &\left.\quad +
    8\log_2\left(\frac{s^{32}+14s^{16}(t^{16}+u^{16})+t^{32}
      +14t^{16}u^{16}+u^{32}}{(s^8+t^8+u^8)^4}\right)\right.\notag\\
      &\left.\quad+\log_2\left(\frac{t^{32}+14t^{16}u^{16}+u^{32}}
    {(t^8+u^8)^4}\right)
    \right]\,.
  \label{magpwres}
\end{align}
These expressions are not unique, due to the ability to recombine
terms using $s + t + u = 0$. However, we note the close
relationship in form between the results, where powers of Mandelstam
invariants in the Yang--Mills theory result are doubled in the gravity result. Again,
this is directly traceable to the KLT relations (alternatively, the
double copy), which multiply together kinematic contributions from
Yang--Mills amplitudes. In figure~\ref{fig:magpow}, we plot the magic
power as a function of the scattering angle. Due to the different
combinations of Mandelstam invariants occurring, and the interplay
between different logarithmic terms, we see that the magic power peaks
closer to $\theta=0$ for gravity, and extends to higher values of
$\theta$ on average for gluons. Consistent with the above remarks, the
peak of the magic power for gravity is lower than that for
Yang--Mills theory. The integrated magic power --- which we can evaluate
numerically --- is also lower:
\begin{align}
  \int_0^{\pi/2} \overline{M_2}(\theta)\Big|_{\rm YM}&=0.245\,,\notag\\
  \int_0^{\pi/2} \overline{M_2}(\theta)\Big|_{\rm grav}&=0.208\,.
\label{magpowint-1}
\end{align}
Thus, the amount of magic generated in gravity is indeed typically
lower than that in a lower-spin theory. Qualitatively similar results
are obtained for higher SREs. For example, in fig.~\ref{fig:M4M10} we
plot the behaviour of $\overline{M_4}$ and $\overline{M_{10}}$,
defined similarly to eq.~(\ref{magicpow}). 

Ref.~\cite{Liu:2025frx} conjectured (with strong numerical
evidence) a theoretical upper bound for the maximum SSRE of a 2-qubit
state:
\begin{equation}
  M_2^{\rm max}=\log\left(\frac{16}{7}\right)\simeq 0.827\,.
  \label{M2max2}
\end{equation}
It is interesting to compare the maximum magic we have found in
eq.~(\ref{M2max}) with this bound, and we find that it is
significantly lower in both Yang--Mills theory and gravity. This also
matches the observation of ref.~\cite{Liu:2025qfl}, which found that
QED is unable to generate the maximum value of magic for most
$2\rightarrow 2$ scattering processes. Nevertheless, that reference
found higher values than observed here, in processes involving
spin-1/2 qubits. Combined with our results above, this suggests that
typical values of magic may decrease as the spin of qubits
increases. This is a very different behaviour to
e.g.\ entanglement, given that maximum entanglement is always possible,
regardless of the spin.

\section{Results for other spins}
\label{sec:SUSY}

To investigate further the issue of how spin affects the typical amount of magic, we can replace the theories considered thus far with their supersymmetric extensions. Supersymmetry adds extra particle content to field theories, where the spins of the additional particles differ from their partners by half integers. Thus, by considering supersymmetric extensions of Yang--Mills theory and gravity, we can repeat our calculations for the scattering of spin-1/2 {\it gluinos}, and spin-3/2 {\it gravitinos}. Whilst the applicability of supersymmetric gauge theory to our own universe may be an open question, the novelty of such theories in our present context is that they provide additional datapoints for what happens when one changes the spin of a qubit.

In order to derive the gluino amplitudes, we can use the supersymmetric Ward identities that relate colour-ordered gluon amplitudes with colour-ordered gluino amplitudes \cite{Grisaru:1977px,Bianchi:2008pu,Dixon:2010ik,Grisaru:1976vm}. Again restricting ourselves to real, $2\rightarrow 2$ kinematics, these identities are given by
\[
\mathcal{A}^{(1/2)}_4[1_{A}^{-}\,2_{B}^{-}\,3_{C}^{+}\,4_{D}^{+}] &= \frac{t\delta_{AC}\delta_{BD} + u\delta_{AD}\delta_{BC}}{s}\mathcal{A}^{(1)}_4[1^-2^-3^+4^+]\\
\mathcal{A}^{(1/2)}_4[1_{A}^{-}\,2_{B}^{+}\,3_{C}^{-}\,4_{D}^{+}] &= \frac{s\delta_{AC}\delta_{BD} + u\delta_{AD}\delta_{BC}}{t}\mathcal{A}^{(1)}_4[1^-2^+3^-4^+]\\
\mathcal{A}^{(1/2)}_4[1_{A}^{-}\,2_{B}^{+}\,3_{C}^{+}\,4_{D}^{-}] &= \frac{s\delta_{AC}\delta_{BD} + t\delta_{AD}\delta_{BC}}{u}\mathcal{A}^{(1)}_4[1^-2^+3^+4^-]\,,
\]
where $A,B,C,D$ are $SU(\cl{N})_R$ flavour indices. The first thing to note is that when there is only a single flavour, as is the case in $\cl{N} = 1$ SUSY, the gluino and gluon amplitudes are identified up to a phase, as can be seen applying $s+t+u = 0$ above along with $A=B=C=D$. Note that the colour ordering does not affect the prefactors, so we can write out the full gluino amplitude in the DDM basis, flipping to incoming momentum in the initial state as before to find

\[
\cl{A}^{(1/2)}[{1}_{A}^{+}\, {2}_{B}^{+}\, ~\rightarrow~ {3}_{C}^{+}\, {4}_{D}^{+}\, ] &= g^2\left(\frac{t\delta_{AC}\delta_{BD} + u\delta_{AD}\delta_{BC}}{s}\right)\left[
    c_{1234}\left(\frac{s}{u}\right)
    +c_{1324}\left(\frac{s^2}{tu}\right)
    \right],\\
\cl{A}^{(1/2)}[{1}_{A}^{+}\, {2}_{B}^{-}\, ~\rightarrow~ {3}_{C}^{-}\, {4}_{D}^{+}\, ] &= g^2\left(\frac{s\delta_{AC}\delta_{BD} + u\delta_{AD}\delta_{BC}}{t}\right)\left[
    c_{1234}\left(\frac{t^2}{su}\right)
    +c_{1324}\left(\frac{t}{u}\right)
    \right],\\
\cl{A}^{(1/2)}[{1}_{A}^{+}\, {2}_{B}^{-}\, ~\rightarrow~ {3}_{C}^{+}\, {4}_{D}^{-}\, ] &= g^2\left(\frac{s\delta_{AC}\delta_{BD} + t\delta_{AD}\delta_{BC}}{u}\right)\left[
    c_{1234}\left(\frac{u}{s}\right)
    +c_{1324}\left(\frac{u}{t}\right)
    \right]\,.
\]
We now take particles $(1,3)$ to share the same flavour, and $(2,4)$ likewise, with the two flavours being distinct. In this setup, the amplitudes reduce to

\[
\cl{A}^{(1/2)}[{1}_{A}^{+}\, {2}_{B}^{+}\, ~\rightarrow~ {3}_{A}^{+}\, {4}_{B}^{+}\, ] &= g^2\left[
    c_{1234}\left(\frac{t}{u}\right)
    +c_{1324}\left(\frac{s}{u}\right)
    \right],\\
\cl{A}^{(1/2)}[{1}_{A}^{+}\, {2}_{B}^{-}\, ~\rightarrow~ {3}_{A}^{-}\, {4}_{B}^{+}\, ] &= g^2\left[
    c_{1234}\left(\frac{t}{u}\right)
    +c_{1324}\left(\frac{s}{u}\right)
    \right],\\
\cl{A}^{(1/2)}[{1}_{A}^{+}\, {2}_{B}^{-}\, ~\rightarrow~ {3}_{A}^{+}\, {4}_{B}^{-}\, ] &= g^2\left[
    c_{1234}
    +c_{1324}\left(\frac{s}{t}\right)
    \right]\,.
\]

The KLT relations in eq. \eqref{M4def} are valid even when the two colour-ordered amplitudes belong to different particle species. Pairing up the four-gluon amplitude with the four-gluino amplitude with the same helicity configuration results in the four-gravitino amplitude. This is an example of the double copy for supersymmetric theories, where such particles are naturally a part of the larger SUSY multiplet.

\begin{figure}[t]
        \centering
        \includegraphics[width=0.70\linewidth]{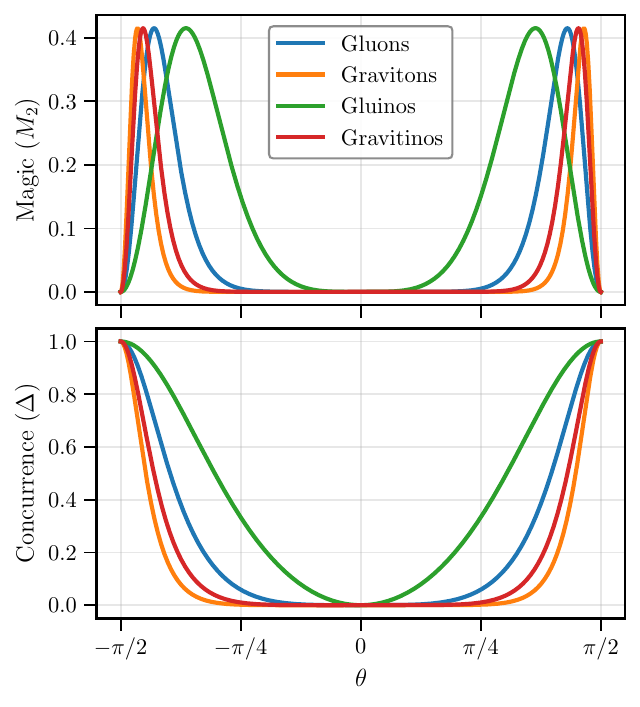}
        \caption
        {(Top) Magic $M_2$ of the final state obtained from initial
          state $|+-\rangle$ for gluons (blue), gravitons (orange), gluinos (green) and gravitinos (red); (bottom) similar, but for the concurrence.}
        \label{fig:gluongravSUSY}
\end{figure}

\begin{figure}[t]
  \begin{center}
    \includegraphics[width=0.9\linewidth]{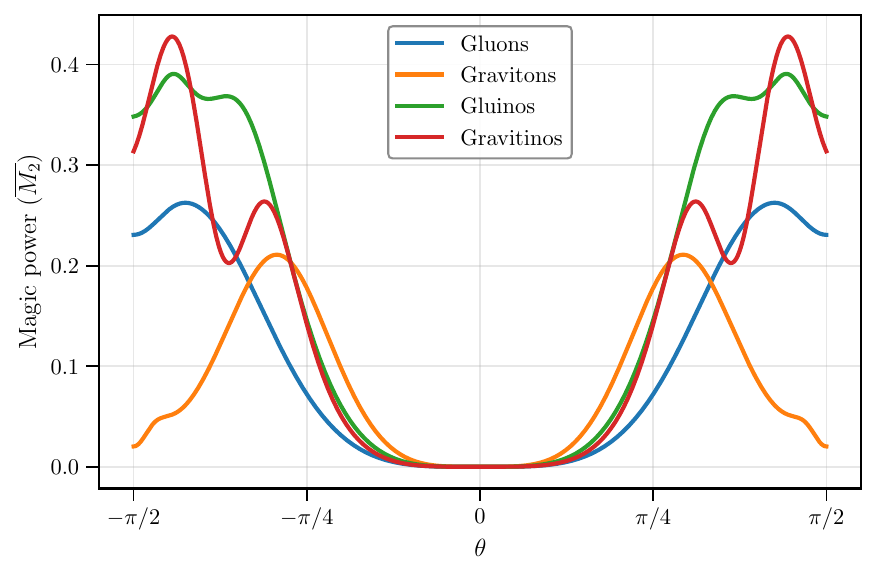}
    \caption{The magic power for gluons (blue), gravitons (orange), gluinos (green) and gravitinos (red) as a function of scattering angle
      $\theta$.}
    \label{fig:magpowSUSY}
  \end{center}
\end{figure}

\begin{figure}[t]
  \begin{center}
    \includegraphics[width=0.9\linewidth]{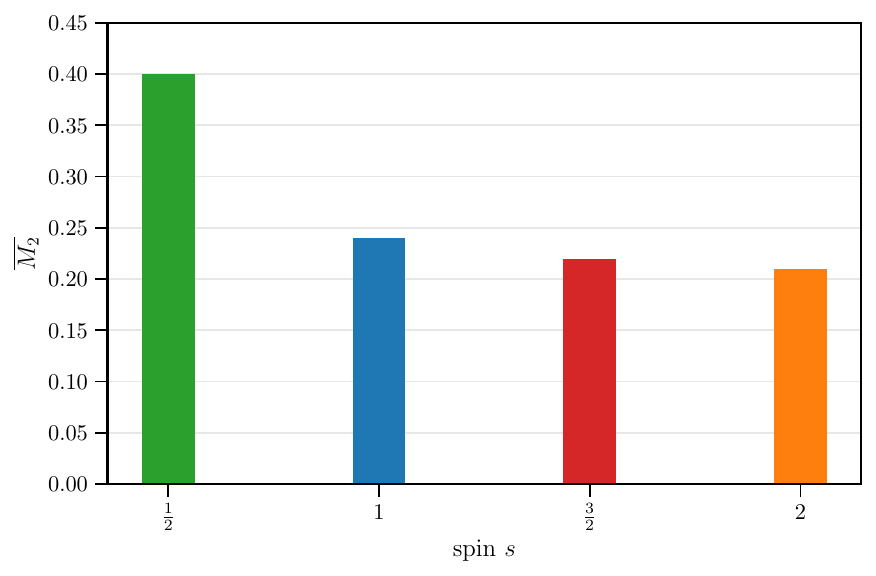}
    \caption{The integrated magic power for gluons (blue), gravitons (orange), gluinos (green) and gravitinos (red), for non-minimal supersymmetry (${\cal N}>1$).}
    \label{fig:magpowint}
  \end{center}
\end{figure}

Picking the first amplitude in the KLT relations to be a colour-ordered gluino amplitude, written in terms of a colour-ordered gluon amplitude, tells us that the gravitino amplitudes satisfy the same Ward identities, and are therefore given by
\[
\mathcal{A}^{(3/2)}[1_{A}^+2_{B}^+~\rightarrow~3_{C}^+4_{D}^+] &= \left(\frac{t\delta_{AC}\delta_{BD} + u\delta_{AD}\delta_{BC}}{s}\right)\mathcal{A}^{(2)}({++\rightarrow ++})\,,\\
\mathcal{A}^{(3/2)}[1_{A}^+2_{B}^-~\rightarrow~3_{C}^-4_{D}^+] &= \left(\frac{s\delta_{AC}\delta_{BD} + u\delta_{AD}\delta_{BC}}{t}\right)\mathcal{A}^{(2)}({+-\rightarrow -+})\,,\\
\mathcal{A}^{(3/2)}[1_{A}^+2_{B}^-~\rightarrow~3_{C}^+4_{D}^-] &= \left(\frac{s\delta_{AC}\delta_{BD} + t\delta_{AD}\delta_{BC}}{u}\right)\mathcal{A}^{(2)}({+-\rightarrow +-})\,.
\]
Again restricting to $A=C$ and $B=D$ this gives
\[
\mathcal{A}^{(3/2)}[1_{A}^+2_{B}^+~\rightarrow~3_{A}^+4_{B}^+] &= -\frac{\kappa^2}{2}\left(\frac{s^2}{u}\right)\,,\\
\mathcal{A}^{(3/2)}[1_{A}^+2_{B}^-~\rightarrow~3_{A}^-4_{B}^+] &= -\frac{\kappa^2}{2}\left(\frac{t^2}{u}\right)\,,\\
\mathcal{A}^{(3/2)}[1_{A}^+2_{B}^-~\rightarrow~3_{A}^+4_{B}^-] &= -\frac{\kappa^2}{2}\left(\frac{u^2}{t}\right)\,.
\]
We will consider the same initial state as in the last section, from which we may compute the magic $M_2$ and concurrence. For gluinos, we find that this is
\begin{equation}
  \Delta(|\psi\rangle_{\rm{gluino}})=\frac{2 t u}{t^2+u^2}\,,\quad
  M_2(|\psi\rangle_{\rm{gluino}})=-\log_2\left(
  \frac{t^{8}+14t^4u^4+u^{8}}{(t^2+u^2)^4}
  \right)\,.
  \label{DeltaM2Gluinos-1}
\end{equation}
whereas for gravitinos we get
\begin{equation}
  \Delta(|\psi\rangle_{\rm{gravitino}})=\frac{2 t^3 u^3}{t^6+u^6}\,,\quad
  M_2(|\psi\rangle_{\rm{gravitino}})=-\log_2\left(
  \frac{t^{24}+14t^{12}u^{12}+u^{24}}{(t^6+u^6)^4}
  \right)\,,
  \label{DeltaM2Gluinos-2}
\end{equation}
and we find that the magic and concurrence for massless particles of any spin $s$ is given by
\[
\Delta(|\psi\rangle_{s})=\frac{2 t^{2s} u^{2s}}{t^{4s}+u^{4s}}\,,\quad
  M_2(|\psi\rangle_{s})=-\log_2\left(
  \frac{t^{16s}+14t^{8s}u^{8s}+u^{16s}}{(t^{4s}+u^{4s})^4}
  \right)\,.
\]
We plot these in figure~\ref{fig:gluongravSUSY}, together with our previous results for gluons and gravitons. As the spin increases from 1/2 to 2, the concentration of the magic and entanglement towards higher scattering angles becomes progressively stronger, thus corroborating the effect seen earlier. 
Similarly as for gluons and gravitons, we can compute the magic power for gluinos and gravitinos, finding
\begin{align}
  \overline{M_2}\Big|_{\rm gluino}&=
  -\frac{1}{15}\left[\log_2 \left(\frac{t^8+14 t^4 u^4+u^8}{\left(t^2+u^2\right)^4}\right)
    +8 \log_2 \left(\frac{16 t^8+28 t^4 u^4+u^8}{\left(2 t^2+u^2\right)^4}\right)\right.\notag\\
    &\left.\quad+
    2 \log_2 \left(\frac{256 t^{16}-512 t^{14} u^2+896 t^{12} u^4-208 t^8 u^8+56 t^4 u^{12}-8 t^2 u^{14}+u^{16}}{\left(4 t^4+u^4\right)^4}\right)
         \right]\,;\notag\\
  \overline{M_2}\Big|_{\rm gravitino}&=
  -\frac{1}{15}\left[
    2\log_2\left(\frac{s^{16}t^8+14s^{8}t^4(t^3+u^4)^4+(t^3+u^3)^8}
    {(s^4t^2+(t^3+u^3)^2)^4}\right)\right.\notag\\
    &\left.\quad
    +2\log_2\left(\frac{s^{16}t^8+14s^{8}(t^4-tu^3)^4+(t^3-u^3)^8}
    {(s^4t^2+(t^3-u^3)^2)^4}\right)\right.\notag\\
    &\left.\quad +
    8 \log_2 \left(\frac{s^{16} t^8+14 s^8 t^4 \left(t^{12}+u^{12}\right)+t^{24}+14 t^{12} u^{12}+u^{24}}{\left(s^4 t^2+t^6+u^6\right)^4}\right)\right.\notag\\
    &\left.\quad+ \log_2 \left(\frac{t^{24}+14 t^{12} u^{12}+u^{24}}{\left(t^6+u^6\right)^4}\right)
    \right]\,.
  \label{magpow2}
\end{align}
The variation of magic power with scattering angle is shown in fig.~\ref{fig:magpowSUSY}, and shows interesting qualitative differences as the spin increases. We note in particular that the interesting structure of peaks and dips in the gravitino results is already present (albeit less pronounced) in the gluino result. Thus, the half-integer spin profiles are more closely related to each other, than to the integer spin results. The integrated magic power for gluinos and gravitinos is found to be
\begin{align}
  \int_0^{\pi/2} \overline{M_2}(\theta)\Big|_{\rm gluinos}&=0.407\,,\notag\\
  \int_0^{\pi/2} \overline{M_2}(\theta)\Big|_{\rm gravitinos}&=0.220\,,
\label{magpowint-0}
\end{align}
and we summarise the comparison with gluons and gravitons in fig.~\ref{fig:magpowint}. In particular, we confirm the pattern of monotonically decreasing magic, as the spin of the qubits increases.

\section{Conclusion}
\label{sec:conclude}

Recent years have seen increasing focus on the quantum property of
{\it magic} (or {\it non-stabliserness}), due to its crucial role in
developing quantum computers with genuine computational advantage, and
that are fault-tolerant. How to produce, enhance and manipulate magic
are open problems, such that case studies that explore these issues in
given physical systems can be useful for generating insights that may
in turn be more general. In this paper, we have considered two-qubit
systems of gluons or gravitons, in Yang--Mills theory and General
Relativity respectively. We have shown that scattering of such
particles indeed generates magic, where the amount of magic depends ---
as in other recent
studies~\cite{White:2024nuc,Aoude:2025jzc,Liu:2025qfl,Fabbrichesi:2025ywl,CMS:2025cim}
--- on the kinematic properties of the final state (i.e.\ the scattering
angle).

Our results provide an interesting case of two closely related
physical systems, that differ by the value of a parameter
(i.e.\ whether, in the non-supersymmetric case, the qubits are spin one or spin two). There is also a
tight theoretical relationship between them, given that amplitudes in
gravity are related to products of gauge theory amplitudes by the KLT
relations~\cite{Kawai:1985xq} (more generally, the double
copy~\cite{Bern:2008qj,Bern:2010ue,Bern:2017yxu,Bern:2010yg}). Thus,
we are able to understand differences in magic between our two
theories in terms of known results in the scattering amplitude
literature, and in particular the generation of higher powers of
Mandelstam invariants in expressions for the magic, in transitioning
from gluons to gravitons. Given that double copy relationships between
different field theories go much wider than traditional gauge theories
and gravity (see e.g.\ ref.~\cite{White:2024pve} for a recent review),
this suggests that our conclusions may well be portable to other
theories. In particular, we see that the typical amount of magic
decreases as the spin of qubits increases, which is also consistent
with previous results for spin-1/2 qubits in
QED~\cite{Liu:2025qfl}. To corroborate this conclusion, we have also calculated results for supersymmetric extensions of Yang--Mills theory and gravity. Such theories contain half-integer spin partners of the gluon and graviton, and show a clear pattern of monotonically decreasing magic as the qubit spin increases. It would be interesting to know if this
conclusion also holds in low energy quantum systems, such as those
being explored in condensed matter physics. We hope that our results
provide a useful contribution to the ongoing investigation of magic
and its uses, and look forward to further work in this area.

\section*{Acknowledgments}

CDW and NM are supported by the UK Science and Technology Facilities Council
(STFC) Consolidated Grant ST/P000754/1 ``String theory, gauge theory
and duality''. MJW is supported by the Australian Research Council
Discovery Project DP220100007, and MJW and JG are both supported by the ARC Centre of Excellence for Dark Matter Particle Physics (CE20010000). ST was supported by the Office of High Energy Physics of the US Department of Energy (DOE) under Grant No.~DE-SC0012567, and by the DOE QuantISED program through the theory consortium “Intersections of QIS and Theoretical Particle Physics” at Fermilab (FNAL 20-17).
ST is additionally supported by the Swiss National Science Foundation project number PZ00P2\_223581, and acknowledges CERN TH Department for hospitality while this research was being carried out. ENVW acknowledges the support he has received for this research through the provision of an Australian Government Research Training Program Scholarship.

\appendix

\section{Kinematics and Ward Identities}
In the centre of mass (CM) frame, we consider the momentum of the incoming particles along the $z$-axis, given by
\[
\begin{aligned}
& p_1^\mu=(E, 0,0, E) \\
& p_2^\mu=(E, 0,0,-E),
\end{aligned}
\]
for the incoming gluons and
\[
\begin{aligned}
& p_3^\mu=(E, E \sin \theta, 0, E \cos \theta) \\
& p_4^\mu=(E,-E \sin \theta, 0,-E \cos \theta),
\end{aligned}
\]
for the outgoing ones, where $\theta$ is the scattering angle. The Mandelstam variables are expressed in terms of $E$ and $\theta$, or $s_{ij} = 2p_i\cdot p_j$, as
\[
s=4 E^2 = s_{12}, \quad t=-4 E^2 \sin ^2\left(\frac{\theta}{2}\right) = -s_{13}, \quad u=-4 E^2 \cos ^2\left(\frac{\theta}{2}\right) = -s_{14}.
\]

In terms of spinor-helicity variables, the momentum can be written as 
\[
p_i^\mu = \bra{i}\sigma^\mu|i] = \lambda_i^a\sigma^{\mu}_{a\dot{a}}\tilde{\lambda}_i^{\dot{a}}
\]
For the kinematics above, the spinors are then real, with $|i] = \ket{i}$, and can be chosen to be
\[
|1\rangle=\sqrt{2 E}\binom{1}{0}, \quad|2\rangle=\sqrt{2 E}\binom{0}{1}, \quad|3\rangle=\sqrt{2 E}\binom{\cos \frac{\theta}{2}}{\sin \frac{\theta}{2}}, \quad|4\rangle=\sqrt{2 E}\binom{\sin \frac{\theta}{2}}{-\cos \frac{\theta}{2}}.
\]
The Mandelstam invariants are then given by $s_{ij} = \braket{ij}[ij] = \braket{ij}^2$, and we can therefore express the spinor brackets in terms of Mandelstam variables via $\braket{ij} = \sqrt{s_{ij}}e^{i\phi_{ij}}$, where $\phi_{ij} = 0$ or $\pi$. For the specific choice of spinors above, we find
\[
\braket{12} = \sqrt{s} = -\braket{34},~~~~~\braket{13} = \sqrt{-t} = -\braket{24},~~~~~\braket{14} = \sqrt{-u} = \braket{23}
\]
and therefore that
\[
\phi_{12} = \phi_{13} = \phi_{14} = \phi_{23} = 0,~~~~~\phi_{34} = \phi_{24} = \pi 
\]
for this choice of kinematics. We can use this to derive the Ward identities used in the main text, where we start from the spinor representation given by \cite{Grisaru:1977px,Bianchi:2008pu,Dixon:2010ik,Grisaru:1976vm}
\[
\mathcal{A}_4^{(1/2)}[{1}_A^- 2_B^- 3_C^+ 4_D^+] = \frac{\braket{13}\braket{24}\delta_{AC}\delta_{BD} - \braket{14}\braket{23}\delta_{AD}\delta_{BC}}{\braket{12}^2}\mathcal{A}^{(1)}_4[1^-2^-3^+4^+]
\]
Using the expressions above, this becomes
\[
\mathcal{A}_4^{(1/2)}[{1}_A^- 2_B^- 3_C^+ 4_D^+] &= \frac{-\sqrt{-t}\sqrt{-t}\delta_{AC}\delta_{BD} - \sqrt{-u}\sqrt{-u}\delta_{AD}\delta_{BC}}{s}\mathcal{A}^{(1)}_4[1^-2^-3^+4^+]\\
&= \frac{t\delta_{AC}\delta_{BD} + u\delta_{AD}\delta_{BC}}{s}\mathcal{A}^{(1)}_4[1^-2^-3^+4^+]
\]
and similarly for the other channels.
\bibliographystyle{utphys}
\bibliography{refs}

\end{document}